\documentclass{aa}

\usepackage{wasysym}

\usepackage{graphicx}
\usepackage{amsmath}   
\usepackage{tabulary}
\usepackage{epsfig}
\usepackage{amssymb}
\usepackage{multirow}
\usepackage{appendix}
\usepackage{natbib}
\usepackage{lineno}
\usepackage{wasysym}
\usepackage{color}         
\usepackage{enumerate}
\usepackage{tikz}
\usepackage{multirow}  

\usepackage{graphicx,url,twoopt,natbib}
\usepackage{txfonts}
\usepackage[pdftitle={The nature of the giant exomoon candidate Kepler-1625\,b-i}, colorlinks = true, breaklinks = true, citecolor = blue, linkcolor = blue, urlcolor = blue, pdfauthor = {Ren\'{e} Heller}]{hyperref}
\bibpunct{(}{)}{;}{a}{}{,}    

\begin{document} 

\title{The nature of the giant exomoon candidate Kepler-1625\,b-i}

\titlerunning{The nature of the giant exomoon candidate Kepler-1625\,b-i}

\author{Ren\'{e} Heller\inst{1}
}

\institute{Max Planck Institute for Solar System Research, Justus-von-Liebig-Weg 3, 37077 G\"ottingen, Germany; \href{mailto:heller@mps.mpg.de}{heller@mps.mpg.de}}

\date{Received 11 August 2017; Accepted 21 November 2017}
 
\abstract{
The recent announcement of a Neptune-sized exomoon candidate around the transiting Jupiter-sized object Kepler-1625\,b could indicate the presence of a hitherto unknown kind of gas giant moon, if confirmed. Three transits of Kepler-1625\,b have been observed, allowing estimates of the radii of both objects. Mass estimates, however, have not been backed up by radial velocity measurements of the host star. Here we investigate possible mass regimes of the transiting system that could produce the observed signatures and study them in the context of moon formation in the solar system, i.e. via impacts, capture, or in-situ accretion. The radius of Kepler-1625\,b suggests it could be anything from a gas giant planet somewhat more massive than Saturn ($0.4\,M_{\rm Jup}$) to a brown dwarf (BD) (up to $75\,M_{\rm Jup}$) or even a very-low-mass star (VLMS) ($112\,M_{\rm Jup}\approx0.11\,M_\odot$). The proposed companion would certainly have a planetary mass. Possible extreme scenarios range from a highly inflated Earth-mass gas satellite to an atmosphere-free water-rock companion of about $180\,M_\oplus$. Furthermore, the planet-moon dynamics during the transits suggest a total system mass of $17.6_{-12.6}^{+19.2}\,M_{\rm Jup}$. A Neptune-mass exomoon around a giant planet or low-mass BD would not be compatible with the common mass scaling relation of the solar system moons about gas giants. The case of a mini-Neptune around a high-mass BD or a VLMS, however, would be located in a similar region of the satellite-to-host mass ratio diagram as Proxima\,b, the TRAPPIST-1 system, and LHS\,1140\,b. The capture of a Neptune-mass object around a $10\,M_{\rm Jup}$ planet during a close binary encounter is possible in principle. The ejected object, however, would have had to be a super-Earth object, raising further questions of how such a system could have formed. In summary, this exomoon candidate is barely compatible with established moon formation theories. If it can be validated as orbiting a super-Jovian planet, then it would pose an exquisite riddle for formation theorists to solve.


}


\maketitle

\section{Introduction}
\label{sec:introduction}

The moons in the solar system serve as tracers of their host planets' formation and evolution. For example, the Earth's spin state, a key factor to our planet's habitability, is likely a result of a giant impact from a Mars-sized object into the proto-Earth \citep{1976LPI.....7..120C}, followed by the tidal interaction of the Earth-Moon binary \citep{1994AJ....108.1943T}. The water contents and internal structures of the Galilean moons around Jupiter have been used to reconstruct the conditions in the accretion disk around Jupiter, in which they supposedly formed \citep{1999SoSyR..33..456M,2002AJ....124.3404C,2015A&A...579L...4H}. The moons around Uranus suggest a collisional tilting scenario for this icy gas giant \citep{2012Icar..219..737M}. It can thus be expected that the discovery of moons around extrasolar planets could give fundamentally new insights into the formation and evolution of exoplanets that cannot be obtained by exoplanet observations alone.

Another fascinating aspect of large moons beyond the solar system is their potential habitability \citep{1987AdSpR...7..125R,1997Natur.385..234W,2006ApJ...648.1196S,2014AsBio..14..798H}. In fact, habitable moons could outnumber habitable planets by far, given their suspected abundance around gas giant planets in the stellar habitable zones \citep{2015A&A...578A..19H}.

Thousands of exoplanets have been found \citep{1995Natur.378..355M,2016ApJ...822...86M}, but no exomoon candidate has unequivocally been confirmed. Candidates have been presented based on a microlensing event \citep{2014ApJ...785..155B}, asymmetries detected in the transit light curves of an exoplanet \citep{2014ApJ...785L..30B}, and based on a single remarkable exoplanet transit in data from \textit{CoRoT} \citep{2015ApJ...805...27L}. Moreover, hints at an exomoon population have been found in the stacked lightcurves of the \textit{Kepler} space telescope \citep{2015ApJ...806...51H}. The most recent and perhaps the most plausible and testable candidate has been announced by \citet{2018AJ....155...36T}. In their search for the moon-induced orbital sampling effect \citep{2014ApJ...787...14H,2016ApJ...820...88H} in exoplanet transit lightcurves from \textit{Kepler}, they found the exomoon candidate Kepler-1625\,b-i.


As the masses of Kepler-1625\,b and its proposed companion remain unknown and the radius of the evolved host star is poorly constrained, the transiting object could indeed be a giant planet. It might also, however, be much larger than Jupiter and, thus, much more massive. Several Jupiter-sized transiting objects from \textit{Kepler} that have been statistically validated but not confirmed through independent methods, such as stellar radial velocity (RV) measurements, later turned out to be very-low-mass stars (VLSMs) rather than planets \citep{2017ApJ...847L..18S}. Here, we report on the plausible masses of Kepler-1625\,b and its satellite candidate and we show to what extent the possible scenarios would be compatible with planet and moon formation scenarios in the solar system.

\section{Methods}

\subsection{System parameters and mass estimates}

\subsubsection{Transit depths and structure models}

Kepler-1625 (KIC\,4760478, KOI-5084), at a distance of $2181_{-581}^{+332}$\,pc, has been classified as an evolved G-type star with a mass of $M_\star=1.079_{-0.138}^{+0.100}\,M_\odot$, a radius of $R_\star=1.793_{-0.488}^{+0.263}\,R_\odot$, an effective temperature of $T_{{\rm eff},\star}=5548_{-72}^{+83}$\,K \citep{2017ApJS..229...30M}, and a Kepler magnitude of $K=13.916$.\footnote{NASA Exoplanet Archive: \href{https://exoplanetarchive.ipac.caltech.edu}{https://exoplanetarchive.ipac.caltech.edu}.} The transiting object Kepler-1625\,b has an orbital period of $287.377314\,{\pm}\,0.002487$\,d \citep{2016ApJ...822...86M}, which translates into an orbital semimajor axis of about 0.87\,AU around the primary star. The next transit can be predicted to occur on 29 October 2017 at $02\,:\,34\,:\,51({\pm}~00:46:18)$\,UT \footnote{The transit epoch of $(2,454,833.0 + 348.83318){\pm0.00729}$\,BJD corresponds to $2455181.834397{\pm0.00729}$\,JD.}, and \citet{2018AJ....155...36T} have secured observations of this transit with the \textit{Hubble Space Telescope}.


The mass of Kepler-1625\,b is unknown. Nevertheless, a range of physically plausible masses can be derived from the observed transit depth and the corresponding radius ratio with respect to the star. The best model fits to the three transit lightcurves from \citet{2018AJ....155...36T} suggest transit depths for Kepler-1625\,b (the primary) and its potential satellite (the secondary) of about $d_{\rm p} = 4.3$\,parts per thousand (ppt) and $d_{\rm s} = 0.38$\,ppt, which translate into a primary radius of $R_{\rm p}=1.18_{-0.32}^{+0.18}\,R_{\rm Jup}$ and a secondary radius of $R_{\rm s}~=~0.35_{-0.10}^{+0.05}\,R_{\rm Jup}~=~0.99_{-0.27}^{+0.15}\,R_{\rm Nep}$. The error bars are dominated by the uncertainties in $R_\star$.

Figure~\ref{fig:M-R} shows a mass-radius curve for non-irradiated substellar objects at an age of 5\,Gyr based on ``COND'' evolution tracks of \citet{2003A&A...402..701B}. Also indicated on the figure are the $1\,\sigma$ confidence range of possible radii of the primary. The surjective nature of the mass-radius relationship prevents a direct radius-to-mass conversion. As a consequence, as long as the mass of Kepler-1625\,b is unknown, for example from long-term RV observations, the radius estimate is compatible with two mass regimes (see the blue bars at the bottom of Fig.~\ref{fig:M-R}). The low-mass regime extends from approximately $0.4\,M_{\rm Jup}$, (roughly 1.3 Saturn masses) to about $40\,M_{\rm Jup}$. The high-mass regime spans from $76\,M_{\rm Jup}$ to about $112\,M_{\rm Jup}\approx0.11M_\odot$. The entire range covers more than two orders of magnitude. Beyond that, the metallicity, age, and rotation state of Kepler-1625\,b might have significant effects on its radius and on the curve shown in Fig.~\ref{fig:M-R}, allowing for an even wider range of plausible masses for a given radius.

The transition between gas giant planets and brown dwarfs (BDs), that is, between objects forming via core accretion that are unable to burn deuterium on the one hand and deuterium-burning objects that form through gravitational collapse on the other hand, is somewhere in the range between $10\,M_{\rm Jup}$ and $25\,M_{\rm Jup}$ \citep{2008A&A...482..315B}. Objects more massive than about $85\,M_{\rm Jup}$ can ignite hydrogen burning to become VLMSs \citep{1963ApJ...137.1121K,1991ARA&A..29..163S}. Based on the available radius estimates alone then, Kepler-1625\,b could be anything from a gas planet to a VLMS.

Similarly, we may derive mass estimates for the proposed companion. The minimum plausible mass for an atmosphere-free object can be estimated by assuming a 50/50 water-rock composition, giving a mass of about $8\,M_\oplus$ \citep{2007ApJ...659.1661F}. Most objects the size of this candidate do have gas envelopes \citep[at least for orbital periods $\lesssim50$\,d;][]{2015ApJ...801...41R}, however, and a substantial atmosphere around the secondary would be likely. In fact, it could be as light as $1\,M_\oplus$, considering the discovery of extremely low-density planets such as those around Kepler-51 \citep{Masuda2014}. On the other end of the plausible mass range, if we consider the upper radius limit and an object composed of a massive core with a low-mass gas envelope akin to Neptune, we estimate a maximum mass of about $20\,M_\oplus$ \citep{2008A&A...482..315B}.

\subsubsection{Transit dynamics}
\label{sec:dynamics}

Given the 287\,d orbit of Kepler-1625\,b, \textit{Kepler} could have observed five transits during its four-year primary mission. Yet, only transits number two (T2), four (T4), and five (T5) of the transit chain were observed. The lightcurves by \citet{2018AJ....155...36T} suggest that T2 starts with the ingress of the proposed satellite, meaning that the satellite would have touched the stellar disk prior to its host. The transit T4 shows the opposite configuration, starting with the primary and ending with the unconfirmed secondary. Then, T5 indicates that the ingress of the moon candidate precedes the ingress of the planet. In contrast to T2, however, the planet leaves the stellar disk first whereas the proposed satellite would still be in transit for an additional $10$\,hr. This suggests a transit geometry in which the moon performs about half an orbit around the planet during the transit, between the two maximum angular deflections as seen from Earth. In fact, the $10$\,hr moon-only part of the transit suggest a sky-projected separation of $17.3\,R_{\rm Jup}$, close to the best fit for the orbital semimajor axis of $a_{\rm ps}=19.1_{-1.9}^{+2.1}\,R_{\rm p}$ \citep{2018AJ....155...36T}. Thus, T5 carries important information about the possible planet-moon orbital period.

 \begin{figure}[t!]
 \centering
  \includegraphics[angle= 0, width=1.\linewidth]{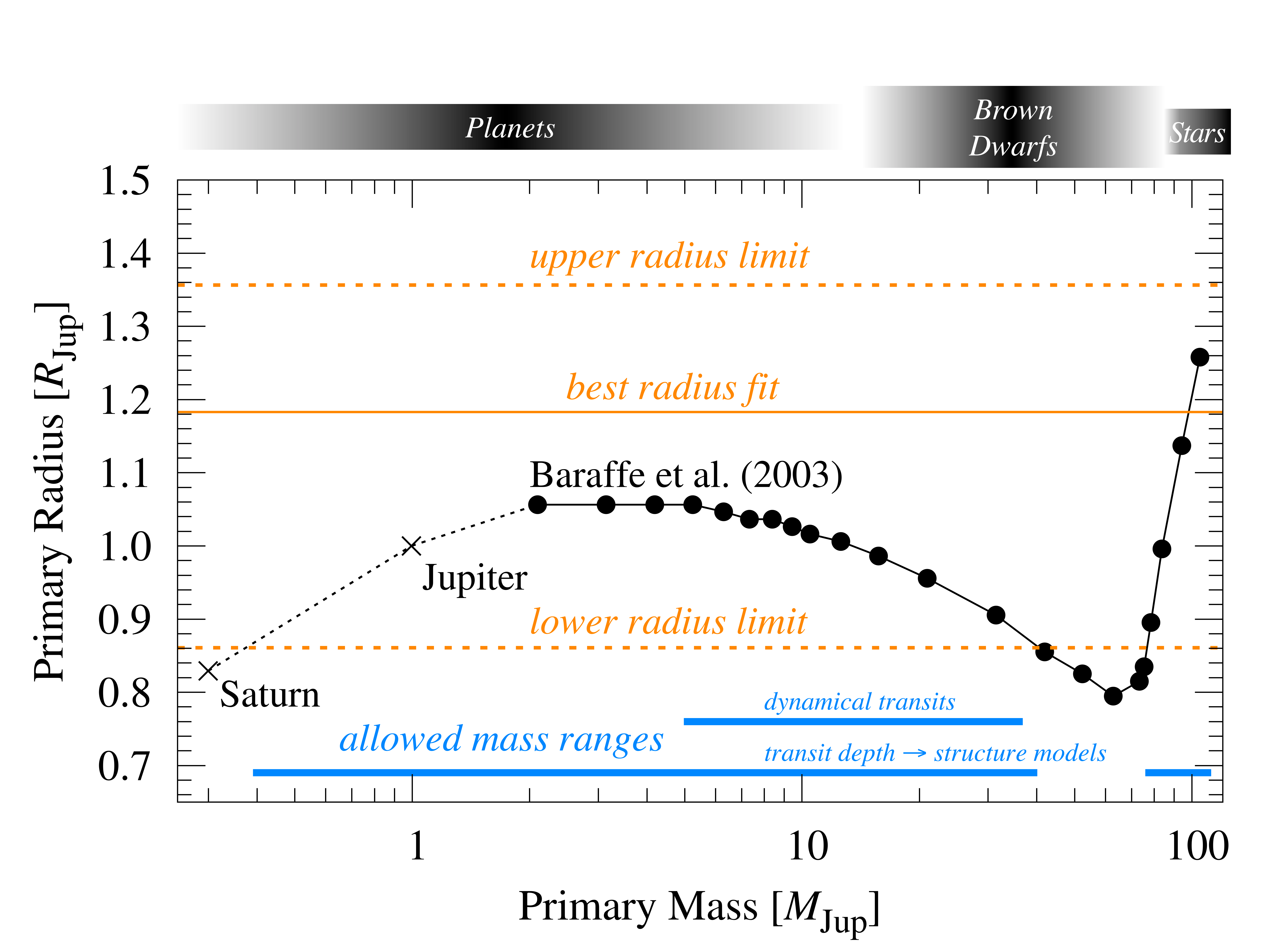}
  \caption{Mass-radius isochrone for substellar objects and VLMSs at an age of 5\,Gyr \citep[][black dots]{2003A&A...402..701B}. Jupiter's and Saturn's positions are indicated with crosses. Horizontal lines indicate the range of possible radii for Kepler-1625\,b estimated from the \citet{2018AJ....155...36T} transit lightcurves and from the uncertainties in the stellar radius. The blue bars at the bottom refer to the mass estimates that are compatible with the isochrone (lower intervals) and as derived from the dynamical signatures in the transits (upper interval).}
  \label{fig:M-R}
 \end{figure}

Assuming that a full orbit would take about twice the time required by the proposed satellite to complete T5 (i.e., its orbital period would be roughly $P_{\rm ps}=72$\,hr), and assuming further an orbital semimajor axis of $a_{\rm ps}=19.1_{-1.9}^{+2.1}\,R_{\rm p}$ \citep{2018AJ....155...36T} together with the uncertainties in the planetary radius, Kepler's third law of motion predicts a barycentric mass of $17.6_{-12.6}^{+19.2}\,M_{\rm Jup}$. This value would be compatible with the ten-Jupiter mass object described by \citet{2018AJ....155...36T}, although it remains unclear how the total mass is shared between the primary and secondary. Despite our neglect of uncertainties in the orbital period, our estimate suggests that proper modeling of the dynamical transit signature can deliver much tighter constraints on the masses of the two bodies than radius estimates alone (see Fig.~\ref{fig:M-R}).

\begin{table*}
\caption{Possible scenarios of the nature of Kepler-1625\,b and its proposed companion.}
\label{tab:scenarios}      
\centering
\begin{tabular}{l | c | c | c | c | c | c}
\hline\hline
Scenario & $R_\star[R_\odot]$ & $R_{\rm p} [R_{\rm Jup}]$ & $R_{\rm s}[R_{\rm Jup}]$ & $M_{\rm p}[M_{\rm Jup}]$ & $M_{\rm s}[M_\oplus]$ & $M_{\rm s}/M_{\rm p}$ \\    
\hline                        
(1aa) {\tiny Saturn-mass gas planet, Earth-mass gas moon} & \multirow{4}{*}{1.305\tablefootmark{(a)}} & \multirow{4}{*}{0.86\tablefootmark{(b)}} & \multirow{4}{*}{0.26\tablefootmark{(b)}} & \multirow{2}{*}{$0.4^{(3)}$} & $1^{(1)}$ & $7.9{\times}10^{-3}$ \\
(1ab) {\tiny Saturn-mass gas planet, Neptune-mass water-rock moon} &                                     &                                   &                                    &                                  & $17^{(2)}$  & $1.3{\times}10^{-1}$ \\
(1ba) {\tiny brown dwarf, Earth-mass gas moon} &                                     &                                   &                                    & \multirow{2}{*}{$75^{(3)}$}   & $1^{(1)}$  & $4.2{\times}10^{-5}$ \\
(1bb) {\tiny brown dwarf, Neptune-mass water-rock moon} &                                     &                                   &                                    &                                  & $17^{(2)}$ & $7.1{\times}10^{-4}$ \\ \hline
(2aa) {\tiny very-low-mass star, mini-Neptune planet} & \multirow{2}{*}{1.793\tablefootmark{(a)}} & \multirow{2}{*}{1.18\tablefootmark{(b)}} & \multirow{2}{*}{0.35\tablefootmark{(b)}} & \multirow{2}{*}{$91^{(3)}$}   & $10^{(4)}$ & $3.5{\times}10^{-4}$ \\
(2ab) {\tiny very-low-mass star, super-Earth water-rock planet} &                                     &                                   &                                    &                                   & $70^{(2)}$& $2.4{\times}10^{-3}$ \\ \hline
(3aa) {\tiny very-low-mass star, Neptune-like planet} & \multirow{2}{*}{2.056\tablefootmark{(a)}} & \multirow{2}{*}{1.36\tablefootmark{(b)}} & \multirow{2}{*}{0.40\tablefootmark{(b)}} & \multirow{2}{*}{$112^{(4)}$} & $20^{(4)}$& $5.6{\times}10^{-4}$ \\
(3ab) {\tiny very-low-mass star, super-Saturn water-rock planet} &                                     &                                   &                                    &                                   & $180^{(2)}$& $5.1{\times}10^{-3}$\\ \hline
(TKS) {\tiny super-Jovian planet, Neptune-like moon} & -- & $1^{(5)}$ & $0.35^{(5)}$ & $10^{(5)}$ & $17^{(5)}$ & $5.4{\times}10^{-3}$ \\ 
\hline
\end{tabular}
\tablefoot{
\tablefoottext{a}{The stellar radius estimates are based on \citet{2017ApJS..229...30M}.}
\tablefoottext{b}{The radii of the transiting primary and proposed secondary were estimated from the lightcurves of \citet{2018AJ....155...36T}.}
The corresponding masses of the objects were estimated using structure models and evolution tracks from the following references.}
\tablebib{(1) \citet{Masuda2014}; (2) \citet{2007ApJ...659.1661F}; (3) \citet{2003A&A...402..701B}; (4) \citet{2008A&A...482..315B}; (5) \citet{2018AJ....155...36T}.}
\end{table*}

 \begin{figure*}[t!]
 \centering
  \includegraphics[angle= 0, width=.613\linewidth]{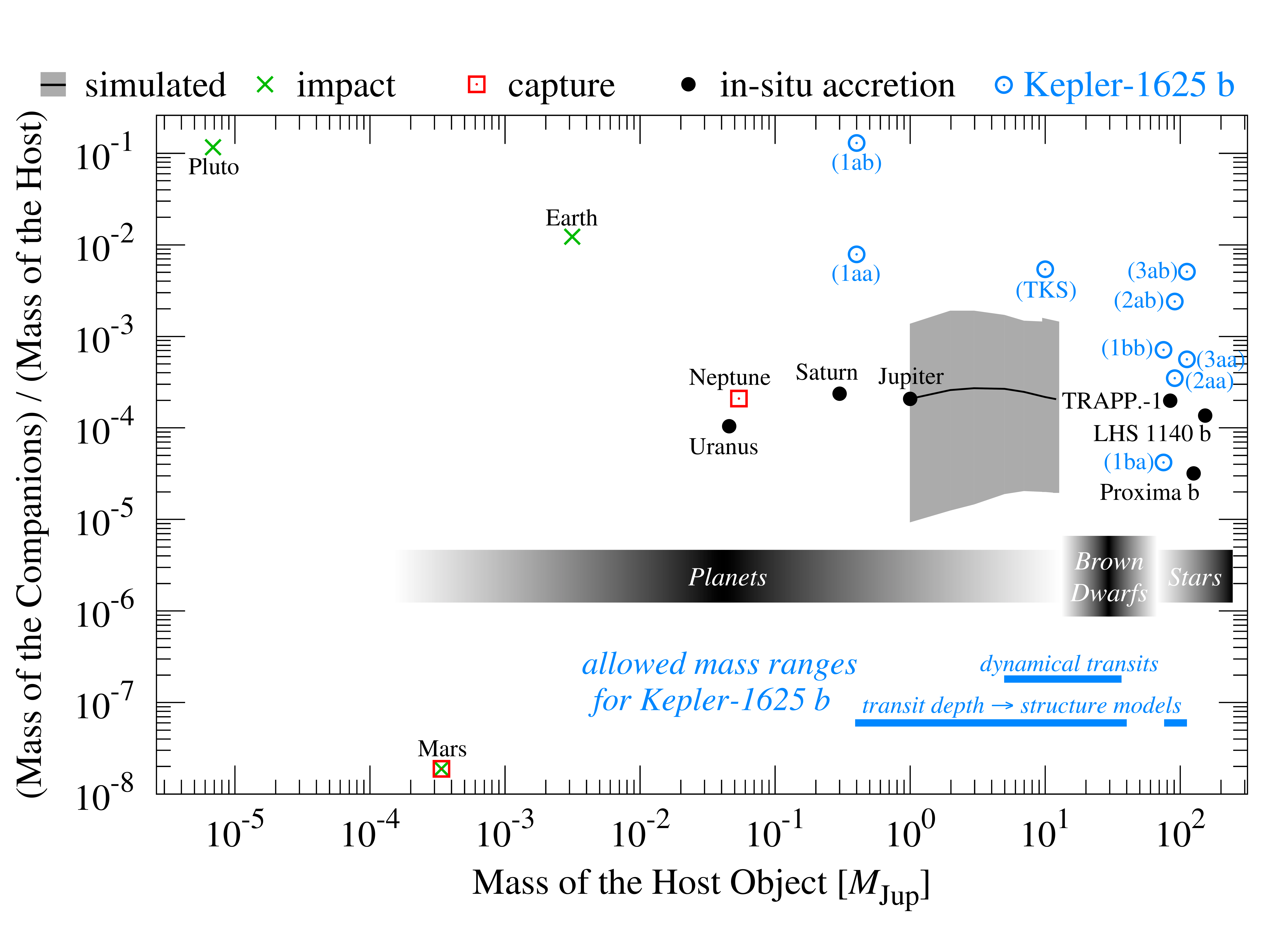}
  \caption{Mass ratios of companions and hosts, i.e., moons around planets and planets around VLMSs. Host masses are shown along the abscissa, mass ratios on the ordinate. Solar system planets with moons are shown with different symbols to indicate the respective formation scenarios of their satellites (see legend above the panel). Three VLMSs with roughly Earth-mass planets (TRAPPIST-1, LHS\,1140, Proxima Centauri) are plotted as examples of formation via accretion in the stellar regime. The solid black line expanding from Jupiter's position signifies simulations of moon formation in the super-Jovian regime, with the gray shaded region referring to uncertainties in the parameterization of the accretion disk. Possible scenarios for the planetary, BD, and VLMS nature of Kepler-1625\,b are indicated with blue open circles (see Table~\ref{tab:scenarios} for details). The plausible mass range for Kepler-1625\,b is shown with a blue line in the lower right corner and is the same as shown in Fig.~\ref{fig:M-R}.}
  \label{fig:ratios}
 \end{figure*}

\subsection{Formation models}

We have investigated the plausible masses of Kepler-1625\,b and its proposed companion in the context of three moon formation scenarios that have been proposed for the solar system moons.

\textit{Impacts}. Moon accretion from the gas and debris disk that forms after giant impacts between planet-sized rocky bodies is the most plausible scenario for the origin of the Earth-Moon system \citep{2012Sci...338.1052C} and for the Pluto-Charon binary \citep{2005Sci...307..546C,2015AJ....150...11W}. A peculiar characteristic of these two systems is in their high satellite-to-host mass ratios of about $1.2{\times}10^{-2}$ for the Earth and $1.2{\times}10^{-1}$ for Pluto.

\textit{In-situ accretion}. In comparison, the masses of the moon systems around the giant planets in the solar system are between $1.0{\times}10^{-4}$ and $2.4{\times}10^{-4}$ times the masses of their host planets. This scaling relation is a natural outcome of satellite formation in a ``gas-starved'' circumplanetary disk model \citep{2006Natur.441..834C} and theory predicts that this relation should extend into the super-Jovian regime, where moons the mass of Mars would form around planets as massive as $10\,M_{\rm Jup}$ \citep{2015ApJ...806..181H}.

\textit{Capture}. The retrograde orbit and the relative mass of Neptune's principal moon Triton cannot plausibly be explained by either of the above scenarios. Instead, \citet{2006Natur.441..192A} proposed that Triton might be the captured remnant of a former binary system that was tidally disrupted during a close encounter with Neptune. The Martian moons Phobos and Deimos have also long been thought to have formed via capture from the asteroid belt. But recent simulations show that they also could have formed in a post-impact accretion disk very much like the proto-lunar disk \citep{Rosenblatt2016}.

To address the question of the origin of Kepler-1625\,b and its potential satellite, we start by calculating the satellite-to-host mass ratios for a range of nominal scenarios for Kepler-1625\,b and its proposed companion. These values shall serve as a first-order estimate of the formation regime which the system could have emerged from.

We then performed numerical calculations of accretion disks around giant planets as per \citet{2015ApJ...806..181H} to predict the satellite-to-host mass ratios in super-Jovian and BD regime, where satellites have not yet been observed. Our disk model is based on the gas-starved disk theory \citep{2006Natur.441..834C,2014SoSyR..48...62M}. While the conventional scenario assumes that the planetary luminosity is negligible for the temperature structure in the disk, we have included viscous heating, accretion heating, and the illumination of the disk by the young giant planet. The evolution of the planet is simulated using pre-computed planet evolution tracks provided by C.~\citet{2013A&A...558A.113M}. We considered seven different giant planets with masses of $1, 2, 3, 5, 7, 10$, and $12$ Jupiter masses. Our model could also include the relatively weak stellar illumination, but we neglected it here to avoid unnecessary complexity.

We tested a range of reasonable disk surface reflectivities, $0.1~{\leq}~k_{\rm s}~{\leq}~0.3$, and assume a constant Planck opacity of $10^{-2}\,{\rm m}^2\,{\rm kg}^{-1}$ throughout the disk. Models in this range of the parameter space have been shown to reproduce the composition and orbital radii of the Galilean moons \citep{2015ApJ...806..181H}. The total amount of solids in the disk is determined by two contributions: (1) the initial solids-to-gas ratio of 1/100, which is compatible with the composition of the interstellar material and in rough agreement with the first direct measurement in a circumstellar disk \citep{2017NatAs...1E.130Z}; (2) by the evolution of the circumplanetary water ice line \citep{2015ApJ...806..181H}, where the surface density of solids (rock and ice) exhibits a jump by a factor of about five as water vapor freezes out \citep{1981PThPS..70...35H}. We measured the amount of solids during the final stages of planetary accretion, when the planet is supposed to open up a gap in the circumstellar disk, at which point the circumplanetary disk is essentially cut from further supply of dust and gas. We assumed that moon formation then halts and compare the final amount of solids to our model of the Jupiter-mass planet. We have not investigated the actual formation of moons from the dust and ice.

To test the plausibility of a capture scenario, we applied the framework of \citet{2013AsBio..13..315W} and calculate the maximum satellite mass that can be captured by Kepler-1625\,b during a close encounter with a binary,

\begin{equation}
M_{\rm cap} < 3 M_{\rm p} \left( \frac{G M_{\rm esc} \pi}{2 b v_{\rm enc} {\Delta}v } \right)^{3/2} - M_{\rm esc} \ ,
\end{equation}

\noindent
where $G$ is the gravitational constant, $M_{\rm esc}$ is the escaping mass, $b$ is the encounter distance, $v_{\rm enc} = \sqrt{v_{\rm esc}^2 + v_{\infty}^2}$ is the encounter velocity, $v_{\infty}$ is the relative velocity of the planet and the incoming binary at infinity, $v_{\rm esc} = \sqrt{2 G M_{\rm p}/b}$ is the escape speed from the planet at the encounter distance $b$, 

\noindent
\begin{equation}
{\Delta}v > \sqrt{ v_{\rm esc}^2 + v_{\infty}^2 } - \sqrt{ G M_{\rm p} \left( \frac{2}{b} - \frac{1}{a} \right) } \ ,
\end{equation}

\noindent
is the velocity change experienced by $M_{\rm cap}$ during capture,

\begin{equation}
a  < \frac{1}{2} \left( 0.5 a_{\rm {\star}p} \left( \frac{M_{\rm p}}{3 M_\star} \right)^{1/3} + b \right)
\end{equation}

\noindent
is the orbital semimajor axis of the captured mass around the planet, and $a_{\rm {\star}p}$ is the semimajor axis of the planet around the star. This set of equations is valid under the assumption that the captured mass would be on a stable orbit as long as its apoapsis were smaller than half the planetary Hill radius.

\section{Results}
\label{sec:results}

Table~\ref{tab:scenarios} summarizes our nominal scenarios for Kepler-1625\,b. Scenarios (1aa) to (1bb) refer to the lower stellar radius estimate, scenarios (2aa) and (2ab) to the nominal stellar radius, and scenarios (3aa) and (3ab) to the maximum stellar radius. In each case, possible values of $R_{\rm p}$ and $R_{\rm s}$ are derived consistently from the respective value of $R_\star$. Scenario (TKS) assumes the preliminary characterization of the transiting system as provided by \citet{2018AJ....155...36T}, except that we explicitly adapted a companion mass equal to that of Neptune although the authors only describe ``a moon roughly the size of Neptune''. The last column in the table shows a list of the corresponding secondary-to-primary mass ratios. Figure~\ref{fig:ratios} illustrates the locations of the said scenarios in a mass ratio diagram. The positions of the solar system planets are indicated as are the various formation scenarios of their satellites (in-situ accretion, impact, or capture). The locations of three examples of planetary systems are included for comparison, namely Proxima\,b \citep{AngladaEscude2016}, the seven planets around TRAPPIST-1 \citep{Gillon2017}, and LHS\,1140\,b \citep{Dittmann2017}.

We find that a Saturn-mass gas planet with either an Earth-mass gas moon (1aa) or a Neptune-mass water-rock moon (1ab) would hardly be compatible with the common scaling law of the satellite masses derived from the gas-starved disk model. The mass ratio would rather be consistent with an impact scenario -- just that the mass of the host object would be much higher than that of any host to this formation scenario in the solar system.

The remaining scenarios, which include a BD orbited by either an Earth-mass gas moon (1ba) or a Neptune-mass water-rock moon (1bb), as well as a VLMS orbited by either an inflated mini-Neptune (2aa), or a water-rock super-Earth (2ab), or a Neptune-like binary planet (3aa), or a water-rock super-Saturn-mass object (3ab) could all be compatible with in-situ formation akin to the formation of super-Earths or mini-Neptunes around VLMSs.

 \begin{figure}[t!]
 \centering
  \includegraphics[angle= 0, width=1.\linewidth]{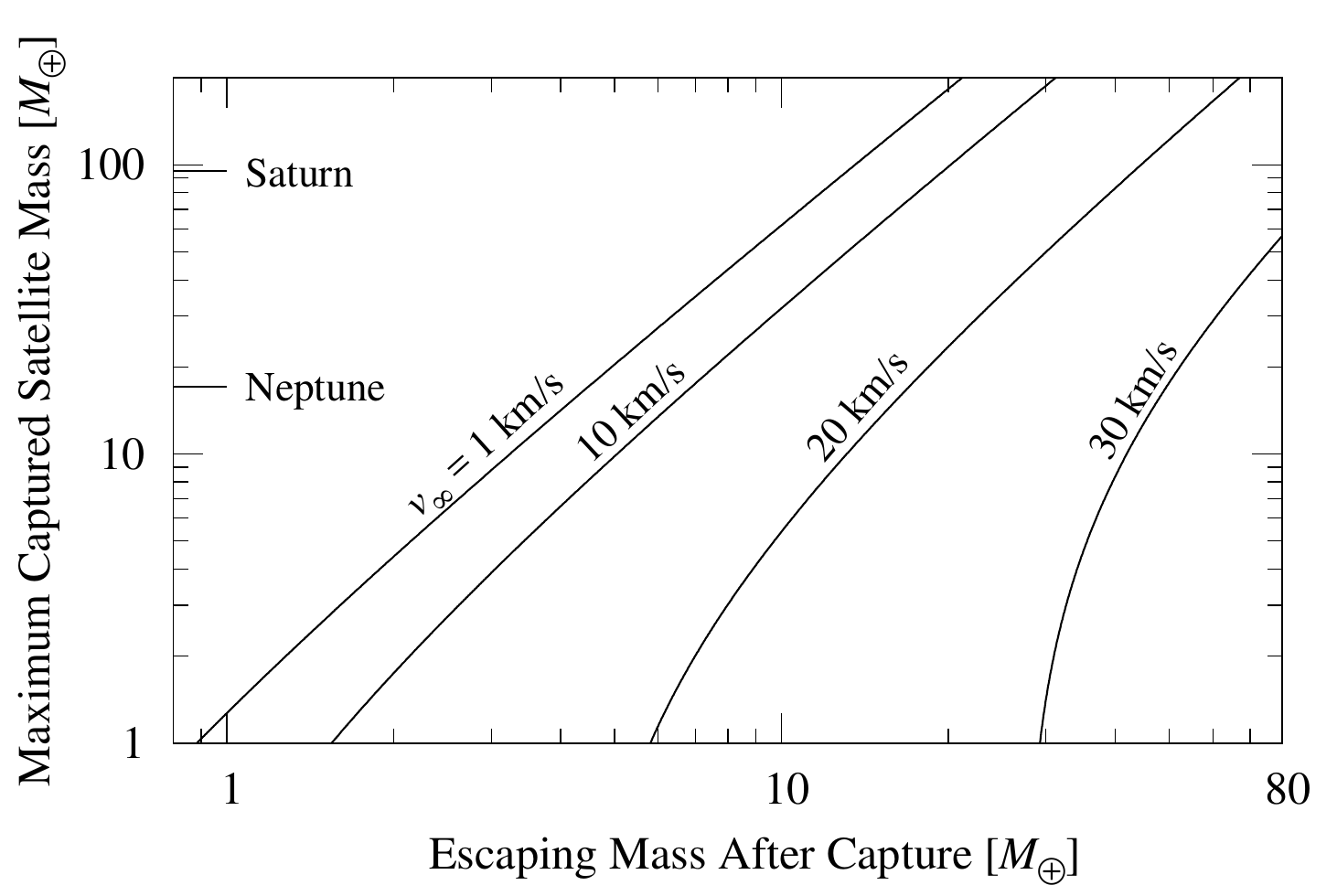}
  \caption{Possible masses for the candidate around Kepler-1625\,b in a capture scenario. The Keplerian speed at the orbit of Kepler-1625\,b is about 30\,km/s and provides a reasonable upper limit on the relative collision speed between it and a possible binary approaching on a close-encounter trajectory.}
  \label{fig:capture}
 \end{figure}

In Fig.~\ref{fig:capture}, we plot the maximum captured mass, that is, the mass of what would now orbit Kepler-1625\,b, over the mass that would have been ejected during a binary encounter with Kepler-1625\,b. We adopted a parameterization that represents the (TKS) scenario, where $M_\star=1.079\,M_\odot$, $a_{\rm {\star}p}=0.87$\,AU, $M_{\rm p}=10\,M_{\rm jup}$ and $b=19.1\,R_{\rm Jup}$. We tested various plausible values for $v_\infty$, ranging from a rather small velocity difference of 1\,km/s to a maximum value of $v_{\rm orb}=30$\,km/s. The latter value reflects the Keplerian orbital speed of Kepler-1625\,b, and crossing orbits would have relative speeds of roughly $e\,{\times}\,v_{\rm orb}$, where $e$ is the heliocentric orbital eccentricity \citep{2006Natur.441..192A}.

We find that the capture of a Neptune-mass object by Kepler-1625\,b is possible at its current orbital location. The escaping object would have needed to be as massive as $5\,M_\oplus$ to $50\,M_\oplus$ for relative velocities at infinity between 1\,km/s and 30\,km/s. For nearly circular heliocentric orbits of Kepler-1625\,b and the former binary, small values of $v_\infty$ would be more likely and prefer the ejection of a super-Earth or mini-Neptune companion from an orbit around the proposed companion to Kepler-1625\,b.

\section{Discussion}
\label{sec:discussion}

The mass estimates derived from the analysis of the lightcurves presented by \citet{2018AJ....155...36T} permit an approximation of the total mass of Kepler-1625\,b and its proposed companion. Photodynamical modeling \citep{Huber2013} accounting for the orbital motion of the transiting system \citep{2011MNRAS.416..689K} as well as the upcoming \textit{Hubble} observations will deliver much more accurate estimates, provided that the proposed companion around Kepler-1625\,b can be validated. If this exomoon candidate is rejected, then the dynamical mass estimate of the planet-moon system (see Sect.~\ref{sec:dynamics}) will naturally become meaningless and a mass-radius relation akin to Fig.~\ref{fig:M-R} will be the only way to estimate the mass of Kepler-1625\,b from photometry alone.

Consistent modeling of the transiting system based on T2, T4, and T5 could allow predictions of the relative orbital geometry of the transiting system during the upcoming transit \citep{2016A&A...591A..67H}, depending on the as yet unpublished uncertainties in $P_{\rm ps}/P_{{\star}p}$ by \citet{2018AJ....155...36T}. If these predictions could be derived and made publicly available prior to the \textit{Hubble} observations of T11 on 29 October 2017, then a confirmation of the proposed planet-moon system in the predicted orbital configuration would lend further credibility to a discovery claim.

The post-capture orbital stability at about 1\,AU around Sun-like stars has been demonstrated for Earth-sized moons around giant planets \citep{2011ApJ...736L..14P}. Most intriguingly, about half of the resulting binaries in these simulations were in a retrograde orbit, akin to Triton around Neptune. Hence, if future observations of Kepler-1625\,b confirm the presence of a companion and if it is possible to determine the sense of orbital motion, then this could be a strong argument for a formation through capture. Measurements of the sense of orbital motion is impossible given even the best available photometric space-based resources nowadays available \citep{2014ApJ...791L..26L,2014ApJ...796L...1H}. Nevertheless, if Kepler-1625\,b turns out to be a BD or VLMS, then its infrared spectrum could be used to determine its RV variation during the transit. Combined with the variation of the tangential motion of Kepler-1625\, and its proposed companion that is available from the lightcurve, this would determine the sense of orbital motion \citep{2017MNRAS.466.4683O}.

The (TKS) scenario, suggesting a Neptune-sized moon in orbit around a $10\,M_{\rm Jup}$ planet, would imply a companion-to-host mass ratio of about $5.4{\times}10^{-3}$, assuming a moon mass equal to that of Neptune. On the one hand, this is just a factor of a few smaller than the relative masses of the Earth-Moon system. On the other hand, this value is more than an order of magnitude larger than the $10^{-4}$ scaling relation established by the solar system giant planets. As a consequence, the satellite-to-host mass ratio does not indicate a preference for either the post-impact formation or in-situ accretion scenarios. In fact, Kepler-1625\,b and its possible Neptune-sized companion seem to be incompatible with both.

If the companion around Kepler-1625\,b can be confirmed and both objects can be validated as gas giant objects, then it would be hard to understand how these two gas planets could possibly have formed through either a giant impact or in-situ accretion at their current orbits around the star. Instead, they might have formed simultaneously from a primordial binary of roughly Earth-mass cores that reached the runaway accretion regime beyond the circumstellar iceline at about 3\,AU \citep{1980ApJ...241..425G}. These cores then may then have started migrating to their contemporary orbits at about 0.87\,AU as they pulled down their gaseous envelopes from the protoplanetary gas disk \citep{1996Natur.380..606L,2015IJAsB..14..201M}.

The case of a VLMS or massive BD with a super-Earth companion would imply a companion mass on the order of $10^{-4}$ times the host object, which reminds us of the proposed universal scaling relation for satellites around gas giant planets. Hence, this scenario might be compatible with a formation akin to the gas-starved disk model for the formation of the Galilean satellites around Jupiter, though in an extremely high mass regime.

\section{Conclusions}
\label{sec:conclusions}

We derived approximate lower and upper limits on the mass of Kepler-1625\,b and its proposed exomoon companion by two independent methods. Firstly, we combined information from the Kepler transit lightcurves and evolution tracks of substellar objects. The radius of Kepler-1625\,b is compatible with objects as light as a $0.4\,M_{\rm Jup}$ planet (similar to Saturn) or as massive as a $0.11\,M_\odot$ star. The satellite candidate would be securely in the planetary mass regime with possible masses ranging between about $1\,M_\oplus$ for an extremely low-density planet akin to Kepler-51\,b or c and about $20\,M_\oplus$ if it was metal-rich and similar in composition to Neptune. These uncertainties are dominated by the uncertainties in the stellar radius. Secondly, we inspected the dynamics of Kepler-1625\,b and its potential companion during the three transits published by \citet{2018AJ....155...36T} and estimate a total mass of $17.6_{-12.6}^{+19.2}\,M_{\rm Jup}$ for the binary. The error bars neglect uncertainties in our knowledge of the planet-moon orbital period, which we estimate from the third published transit to be about $72$\,hr.

We conclude that if the proposed companion around Kepler-1625\,b is real, then the host is most likely a super-Jovian object. In fact, a BD would also be compatible with both the mass-radius relationship for substellar objects and with the dynamical transit signatures shown in the lightcurves by \citet{2018AJ....155...36T}. If the satellite candidate can be confirmed, then dynamical modeling of the transits can deliver even better mass estimates of this transiting planet-moon system irrespective of stellar RVs.

Our comparison of the characteristics of proposed exomoon candidate around Kepler-1625\,b with those of the moon systems in the solar system reveal that a super-Jovian planet with a Neptune-sized moon would hardly be compatible with conventional moon formation models, for example, after a giant impact or via in-situ formation in the accretion disk around a gas giant primary. We also investigated formation through an encounter between Kepler-1625\,b and a planetary binary system, which would have resulted in the capture of what has provisionally been dubbed Kepler-1625\,b-i and the ejection of its former companion. Although such a capture is indeed possible at Kepler-1625\,b's orbital distance of 0.87\,AU from the star, the ejected object would have had a mass of a mini-Neptune itself. And so this raises the question how this ejected mini-Neptune would have formed around the Neptune-sized object that is now in orbit around Kepler-1625\,b in the first place. If the proposed exomoon can be validated, then measurements of the binary's sense of orbital motion might give further evidence against or in favor of the capture scenario.


The upcoming \textit{Hubble} transit observations could potentially allow a validation or rejection of the proposed exomoon candidate. If the moon is real, then dynamical transit modeling will allow precise mass measurements of the planet-moon system. If the moon signature turns out to be a ghost in the detrended {\it Kepler} data of \citet{2018AJ....155...36T}, however, then the transit photometry will not enable a distinction between a single transiting giant planet and a VLMS. Stellar spectroscopy will then be required to better constrain the stellar radius, and thus the radius of the transiting object, and to determine the mass of Kepler-1625\,b or Kepler-1625\,B, as the case may be.

\begin{acknowledgements}
The author thanks Laurent Gizon, Matthias Ammler-von Eiff, Michael Hippke, and Vera Dobos for helpful discussions and feedback on this manuscript. This work was supported in part by the German space agency (Deutsches Zentrum f\"ur Luft- und Raumfahrt) under PLATO Data Center grant 50OO1501. This work made use of NASA's ADS Bibliographic Services. This research has made use of the NASA Exoplanet Archive, which is operated by the California Institute of Technology, under contract with the National Aeronautics and Space Administration under the Exoplanet Exploration Program.
\end{acknowledgements}


\bibliographystyle{aa} 
\bibliography{ms}

\end{document}